\documentclass[11pt]{article}
\usepackage{hyperref}
\usepackage{natbib}
\usepackage{bm}
\usepackage{amsmath}
\usepackage{amsthm}
\usepackage{amsfonts}
\usepackage{booktabs}
\usepackage[justification=centering]{caption}
\usepackage{mathtools}
\usepackage{graphicx}
\usepackage{subcaption}
\graphicspath{{Figures/}}
\usepackage{float}
\theoremstyle{definition}

\newcommand{\me}{\mathrm{e}}
\newcommand{\EX}{\mathbb{E}}
\def\*#1{\mathbf{#1}}
\newcommand{\RN}[1]{%
  \textup{\uppercase\expandafter{\romannumeral#1}}%
}

\DeclareMathOperator*{\argmin}{arg\,min}
\DeclarePairedDelimiter\abs{\lvert}{\rvert}
\title{Minimum message length inference of the Poisson and geometric models using heavy-tailed prior distributions}
\author{Chi Kuen Wong, Enes Makalic, Daniel F. Schmidt}
\begin{document}
\maketitle
\begin{abstract}
Minimum message length is a general Bayesian principle for model selection and parameter estimation that is based on information theory. This paper applies the minimum message length principle to a small-sample model selection problem involving Poisson and geometric data models. Since MML is a Bayesian principle, it requires prior distributions for all model parameters. We introduce three candidate prior distributions for the model parameters with both light- and heavy-tails. The performance of the MML methods is compared with objective Bayesian inference and minimum description length techniques based on the normalized maximum likelihood code. Simulations show that our MML approach with a heavy-tail prior distribution performs well in all tests.
\end{abstract}
%
%\begin{keyword}
%minimum message length \sep Poisson \sep model selection \sep geometric
%\end{keyword}
\section{Introduction}
Model selection is a fundamental task in statistics. With today's computing technology, fitting a model to data using standard statistical software is often a straightforward task. But, how should we decide if one model is better than another? One major concept in model selection is the principle of parsimony, closely related to the Occam's razor, which states that ``more things should not be used than are necessary''. Many model selection techniques implicitly employ this principle. 

In this paper, we investigate a specific model selection problem involving the Poisson and geometric probability distributions. Suppose we are given some data $\*x = (x_1, \ldots, x_n) \in \mathbb{N}^n$ with the hypothesis that the data is generated either from a Poisson or a geometric model. Our task is to decide which of these two models best explains the data. The Poisson and geometric models are both single parameter models. The probability mass functions, denoted by $f_{P}$ for the Poisson and $f_{G}$ for the geometric model, are
\begin{align*}
	f_P(x|\lambda) &= \frac{\lambda^x \me^{-\lambda}}{\Gamma(x+1)}  \,, \hspace{0.2mm} \quad (\lambda > 0), \\
	f_G(x|p) &= (1-p)^x p \,, \quad (0 < p \leq 1), 
\end{align*}
for $x\in\{0,1,2,\dots\}$, where $\Gamma(\cdot)$ is the gamma function. We use the above parameterization of the geometric distribution since it has the same support as the Poisson distribution.

We now briefly list some important statistical properties of the Poisson and geometric distributions that will be used throughout this paper. Given a sample $\*x=(x_1,\dots,x_n)$ with $n$ independent and identically distributed observations, the likelihood functions for the Poisson and geometric models are:
\begin{align*}
	%\label{eq:like.pois}
	f_P(\*x|\lambda) &= \prod_{i=1}^{n} \frac{\lambda^{x_{i}} \me^{-\lambda}}{\Gamma(x_{i}+1)} = \frac{\lambda^{s}\me^{-n\lambda}}{\prod_{i=1}^{n} \Gamma(x_{i}+1)}\,, \\
	%
	%\label{eq:like.geom}
	f_G(\*x|p) &=  \prod_{i=1}^{n} (1-p)^{x_i}p = (1-p)^{s}p^{n} \,, 
\end{align*}
where $s(\*x)\equiv s =\sum_{i=1}^n x_i$ is the minimal sufficient statistic. The corresponding negative log-likelihood functions are: 
\begin{align}
	\label{eq:loglike.pois}
	l_P(\*x|\lambda) & = -s \ln(\lambda) + \lambda n + \sum_{i=1}^n \ln \Gamma(x_{i}+1) \,, \\
	\label{eq:loglike.geom}
	l_G(\*x|p) &= -s \ln(1-p) - n \ln p  \,. 
\end{align}
The maximum likelihood estimators (MLE) for the Poisson and geometric distributions, found by minimizing (\ref{eq:loglike.pois}) and (\ref{eq:loglike.geom}) respectively, are 
\begin{align*}
	\hat{\lambda}(\*x) &= \frac{s}{n}\,, \\
	\hat{p}(\*x) &=  \frac{n}{n + s}\,.
\end{align*}
Lastly, the Fisher information for the Poisson and geometric models is
\begin{align}
	\label{eq:fish.pois}
	F_P(\lambda) &= \frac{n}{\lambda} \,, \\ 
	\label{eq:fish.geom}
	F_G(p) &= \frac{n}{p^2(1-p)} \,.
\end{align}

This paper builds upon the work of \cite{de2006empirical}, who used the minimum description length (MDL) principle \citep{rissanen1998stochastic, rissanen2007information}, based on the normalized maximum likelihood (NML) code, and objective Bayesian approaches to tackle the above problem. Note that \cite{de2006empirical} used an alternative parameterization for the geometric model; they re-parmeterized the geometric distribution by its mean $\mu=(1-p)/p$. The probability mass function under this parameterization is
\begin{equation}
	\label{eq:geom.para.mean}
	f_G(x|\mu) = \left(\frac{\mu}{1+\mu}\right)^{x} \left( \frac{1}{1+\mu} \right) \,, \quad (\mu > 0).
\end{equation}
The negative log-likelihood function is
\begin{equation}
	\label{eq:loglike.geom2}
	l_G(\*x|\mu) = - s \ln(\mu) + \left(s + n\right)\ln(1+\mu) \,,
\end{equation}
the Fisher information is
\begin{equation}
	\label{eq:fish.geom2}
	F_G(\mu) = \frac{n}{\mu(1+\mu)} \,,
\end{equation}
and the MLE for $\mu$ is
\begin{equation*}
	\hat{\mu}(\*x) = \frac{s}{n} \,.
\end{equation*}

The aim of this paper is to introduce the minimum message length (MML) principle \citep{wallace2005statistical, wallace1968information, wallace1975invariant, wallace1987estimation}, which is closely related to MDL, and apply it to the model selection problem involving the Poisson and geometric models. 

Section \ref{sec:mdl} and Section \ref{sec:mdl.pois.geom} provide an overview of MDL and model selection techniques examined by \cite{de2006empirical}. In Section \ref{sec:mml}, we describe the MML approach, which is inherently Bayesian, and propose three different prior distributions for the model parameters. Section \ref{sec:sim} presents our simulation results comparing the performance of our MML technique with the MDL approaches and the objective Bayesian inference methods. The conclusion is given in Section \ref{sec:con}.
\section{Minimum Description Length}
\label{sec:mdl}
The insight that knowledge can be gained by compressing data is the foundation of the MDL principle. In the MDL framework, the best model is defined as the one that compresses the data as much as possible. Let $\mathcal{X}^n$ be an $n$-dimensional sample space, $\*x \in \mathcal{X}^n$ be a data sequence, and $\mathcal{C}$ be the set of all candidate binary codes which will be used to encode (i.e., represent) the data sequence. We define a binary code to be a function that maps every possible data sequence to some string $C \in \cup_{m \geq 1} \left\{0,1\right\}^m$, where $\left\{0,1\right\}^m$ is the set of binary strings with $m$ digits. The length of any code $C$ is equal to the number of digits in the code. In this paper, we do not focus on the encoding process (i.e., how to represent a data sequence using a binary string), and instead examine approaches for computing the codelengths of the binary data strings.  

In the MDL framework, we must decide which code to use before a data sequence is observed. Since we are only interested in the codelength, we can represent any code $C \in \mathcal{C}$ by the corresponding codelength function $l_C(\*x)$. From Shannon's theory of information \citep{shannon1948mathematical}, the relation between a probability distribution $P(\*x)$ and the corresponding codelength function is
\begin{equation}
\label{eq:Shannon}
	l_C(\*x) = -\log_2 P(\*x) \,.
\end{equation}
This means that we use longer codewords to encode rarer data sequences and shorter codewords for data sequences that are more common. The unit of a codelength depends on the base of the logarithm in (\ref{eq:Shannon}). If the measurement of information is based on base-$2$ logarithms the unit is called a bit (binary digit), and if the measurement is based on natural logarithms, the unit is called a nit or nat. 

Ideally, we want to find a code $C$ such that no matter what data $\*x$ is observed, the codelength $l_C(\*x)$ is minimum among all possible candidate codes. Unfortunately, this code, which is usually referred to as the ideal code, does not exist~\citep{grnwald2005advances}. However, it can be shown that codes exist such that for any data $\*x$, they perform almost as well as the ideal code. In other words, there exists a code $U$ with codelength function $l_U$ such that for all $\*x \in \mathcal{X}^n$: 
\begin{equation*}
	l_U(\*x) \leq \inf_{C \in \mathcal{C}} l_C(\*x) + K \,,
\end{equation*}
where $K$ is some constant that does not depend on $n$. Such codes are referred to as universal codes, and a probability distribution corresponding to a universal code is called a universal model.

Formally, suppose $\mathcal{M}$ is family of distributions characterized by the density function $p(x | \theta)$, that is, $\mathcal{M}=\{p(x|\theta_1),p(x|\theta_2),\dots,p(x|\theta_m)\}$, where the number of distributions in $\mathcal{M}$ is finite. A model with density $\bar{p}$ is called a universal model relative to $\mathcal{M}$, if for all $\theta$ and $\*x$: 
\begin{equation*}
	-\ln \bar{p}(\*x) \leq -\ln p(\*x|\theta) + K \,.
\end{equation*} 
%
%%!! MORE TEXT %% 
This means that a code for data $\*x$ based on a universal model is at most $K$ nits longer compared to a code based on any of the models defined in the set of distributions $\mathcal{M}$, where $K$ is independent of sample size $n$.

To define a universal model within the MDL framework we require the concept of coding regret. The regret of using a particular model characterized by the density $\bar{p}$, relative to a family of distributions $\mathcal{M}$, is 
\begin{equation}
\label{eq:regreg}
R(\*x,\bar{p}) = -\ln\bar{p}(\*x) + \ln p(\*x|\hat{\theta}(\*x)) \,,
\end{equation} 
where $\hat{\theta}(\*x)$ is the MLE. The regret is the additional codelength required to encode the data compared to the best-fitting model in $\mathcal{M}$. Note that the best-fitting model is not decodable as it requires knowledge of the data. The MDL principle seeks a universal model $\bar{p}$ such that the regret is at its minimum for the worst case data sequence:
\begin{equation}
\label{eq:minimax.regret}
	\min_{\bar{p}\in P} \left\{\max_{\*x\in \mathcal{X}}\left[-\ln\bar{p}(\*x) + \ln p(\*x|\hat{\theta}(\*x))\right]\right\} \,,
\end{equation}
where $P$ is the set of measurable probability distributions. \cite{shtar1987universal} found that the solution to (\ref{eq:minimax.regret}) is the normalized maximum likelihood (NML) distribution
\begin{equation*}
	p_\textrm{\tiny{NML}}(\*x) = \frac{p\left(\*x|\hat{\theta}(\*x)\right)}{\sum_{{\bf y}} p\left({\bf y}|\hat{\theta}({\bf y})\right)} \,.
\end{equation*}
The codelength of data $\*x$ coded using the NML distribution is
\begin{equation}
	\label{eq:nml.code}
	-\ln p_\textrm{\tiny{NML}}(\*x) = -\ln p\left(\*x|\hat{\theta}(\*x)\right) + \ln \sum_{{\bf y}} p\left({\bf y}|\hat{\theta}({\bf y})\right) \,,
\end{equation}
%
%%!! MORE TEXT %%
where the last term is known as the parametric complexity. This can be interpreted as a measure of complexity of a model class, and measures how well the model class fits random data sequences. The greater the parametric complexity of $\mathcal{M}$, the greater the number data sequences that can be fitted well using the models from $\mathcal{M}$. The parametric complexity can also be interpreted as the minimax regret relative to $\mathcal{M}$, which is the worst case additional codelength required to encode data compared to the best model in $\mathcal{M}$. 

%One possible difficulty in computing the NML codelength is that the parametric complexity term may become infinite. In this case, 
In many model classes, the exact parametric complexity is unavailable in closed form and is difficult to compute numerically. Consequently, researchers resort to approximations and the most popular approximation to the parametric complexity term is \citep{rissanen1996fisher, takeuchi1997asymptotically, takeuchi1998asymptotically}:
\begin{equation}
	\label{eq:pc.approx}
	\ln\sum_{\*y} p(\*y|\hat{\theta}(\*y)) = \frac{k}{2}\ln\frac{n}{2\pi} + \ln\int_{\Theta}\sqrt{|F_1(\theta)|} d\theta \, + o(1) \,,
\end{equation}
where $k > 0$ is the number of model parameters and $|F_1(\theta)|$ is the determinant of the Fisher information matrix for a single data point. For more information on MDL and its properties the reader is directed to~\cite{grunwald2007minimum, myung2006model}.
\section{MDL approaches in Poisson and geometric models}
\label{sec:mdl.pois.geom}
One possible difficulty in computing the NML codelength (\ref{eq:nml.code}) is that the parametric complexity term may be infinite. In fact, in the case of the Poisson and geometric models, not only is the parametric complexity term infinite, but the approximation also diverges -- the integral in (\ref{eq:pc.approx}) is not finite for either model.  Several approaches to overcome this problem have been examined by \cite{de2006empirical}, including the Bayesian information criterion (BIC), the restricted approximate normalized maximum likelihood (ANML), the two-part ANML, the objective Bayesian code, and the plug-in predictive code. Of these methods, \cite{de2006empirical} showed that the plug-in predictive code had poor performance in simulations and is therefore not considered in the remainder of this paper. 

We will now review the other four approaches considered by \cite{de2006empirical} to computing the (approximate) parametric complexity, namely the BIC, ANML, two-part ANML and the objective Bayesian code. We use $I_{P}$ and $I_{G}$ to denote the codelength for the Poisson and geometric models, respectively. The functions $l_P$ and $l_G$ are the corresponding negative log-likelihood functions, given in (\ref{eq:loglike.pois}) and (\ref{eq:loglike.geom2}), $F_P$ and $F_G$ are the Fisher information terms, given in (\ref{eq:fish.pois}) and (\ref{eq:fish.geom2}), and $\hat{\lambda}$ and $\hat{\mu}$ are the MLEs, which are both equal to $\sum_{i=1}^n x_i/n \equiv s/n$. Also note that in this section, the geometric model uses the mean parameterization as in (\ref{eq:geom.para.mean}) and the codelengths are measured in nits. 

\paragraph{BIC}
One simple way to resolve the problem of infinite parametric complexity is to drop the infinite integral term in (\ref{eq:pc.approx}). In this case, the NML codelength becomes exactly equal to the popular BIC~\citep{schwarz1978estimating,rissanen1978modeling}:
\begin{align*}
	I_{P}(\*x) &= l_{P}(\*x|\hat{\lambda}) + \frac{1}{2} \ln n \,, \\
	I_{G}(\*x) &= l_{G}(\*x|\hat{\mu}) + \frac{1}{2} \ln n \,,
\end{align*}
where the last term in both codelength formulas is the same as both models have only one free parameter. As such, comparing the BIC values for the two models is equivalent to comparing the negative log-likelihoods of the Poisson and geometric models evaluated at the MLE.

\paragraph{Restricted ANML}
In this approach, instead of computing the (infinite) integral in (\ref{eq:pc.approx}) over the entire parameter space $\mu \in \mathbb{R}^+$, we evaluate this integral over the restricted region $\mu \in (0, \mu^*]$: 
\begin{align*}
	%\label{eq:rnml.pois}
	I_{P}(\*x) &= l_{P}(\*x|\hat{\lambda}) + \frac{1}{2} \ln \frac{n}{2\pi} + \ln\int_0^{\mu^*} u^{-\frac{1}{2}}du \,, \\
	%
	%\label{eq:rnml.geom}
	I_{G}(\*x) &= l_{G}(\*x|\hat{\mu}) + \frac{1}{2} \ln \frac{n}{2\pi} + \ln\int_0^{\mu^*}\frac{du}{\sqrt{u(u+1)}} \,.
\end{align*}
%
%The last terms in (\ref{eq:rnml.pois}) and (\ref{eq:rnml.geom}) are approximations of the parametric complexity for the original NML code (\ref{eq:nml.code}). 
We must specify the value of $\mu^*$ to compute the approximate codelengths. As the choice of the parameter region is arbitrary, the codelength will be different for different values of $\mu^*$ which may lead to good model selection performance on one data set, but bad performance in another.
%
% Following \cite{de2006empirical}, we have used $\mu^*=\left\{10,100,1000\right\}$ in our simulation in Section \ref{sec:sim}. 
%
%
%
\paragraph{Two-part ANML}
In the two-part ANML approach, we first encode an integer $b = \lceil\ln_{2}\hat{\mu}\rceil$, where $\hat{\mu}$ is the MLE of $\mu$. We then encode the data using the restricted ANML approach on the range $(2^{b-1},{2^b}]$, resulting in the following codelengths:
\begin{align*}
	%\label{eq:nml.two.part}
	I_P(\*x) &= l_{P}(\*x|\hat{\lambda}) + \frac{1}{2} \ln \frac{n}{2\pi} + \ln \left( \int_{2^{b-1}}^{2^b} u^{-\frac{1}{2}}du \right) + l^*(b) \,, \\
	I_G(\*x) &= l_{G}(\*x|\hat{\mu}) + \frac{1}{2} \ln \frac{n}{2\pi} + \ln\left( \int_{2^{b-1}}^{2^b}\frac{du}{\sqrt{u(u+1)}} \right) + l^*(b) \,.
\end{align*}
Here $l^*(b) = \left[\log_2^*(b) + \log_2(2.865604)\right]\times\ln 2$ and $\log_2^*(b)$ is the log-star code for the integer $b$~\citep{rissanen2007information,rissanen1983universal}, defined as:
\begin{equation*}
	\log_2^*(b) = \log_2(b)+\log_2\log_2(b) + \log_2\log_2\log_2(b) + \dots 
\end{equation*}
where the last term in the sum is the last positive term. Although in this approach we do not have to select an arbitrary parameter region as per the ANML code, the restricted ANML codelength is no longer minimax optimal. %%!! WHAT IS MINIMAX OPTIMAL? NEED TO DEFINE THIS IN THE MDL SECTION SOMEHWERE AFTER NML DISCUSSION %%
\paragraph{Objective Bayesian code}
\cite{de2006empirical} also examined the objective Bayesian and the approximate objective Bayesian approach. The idea here is to first use the Jeffreys prior distribution~\citep{jeffreys1946invariant} for the unknown parameter and compute the posterior distribution of the parameter using only the first observation $x_1$. We then use this posterior distribution as a prior distribution for the remaining data $\*x_2^n = (x_2,\dots,x_n)$ in a Bayesian framework. The codelength of the objective Bayesian code is obtained by taking the negative log of the marginal likelihood, which is defined as
\begin{equation}
	\label{eq:marginal}
	m(\*x) = \int_{\Theta} \pi(\theta) f(\*x|\theta) d\theta \,,
\end{equation}
where $\pi(\cdot)$ is the prior distribution for $\theta$ and $f(\*x|\theta)$ is the likelihood function. The objective Bayesian codelengths for the Poisson and geometric models are
\begin{align*}
I_{P}(\*x_2^n|x_1) &= \ln\Gamma\left(x_1+\tfrac{1}{2}\right) -\ln\Gamma\left(s+\tfrac{1}{2}\right) + \left(s+\tfrac{1}{2}\right)\ln(n) + \sum_{i=2}^{n} \ln\Gamma(x_i +1) \,, \\
I_{G}(\*x_2^n|x_1) &= -\ln\left(x_1+\tfrac{1}{2}\right) - \ln\Gamma\left(s+\tfrac{1}{2}\right) - \ln\Gamma(n) + \ln\Gamma\left(n+s+\tfrac{1}{2}\right) \,,
\end{align*}
where $x_1$ is the first observation of the data $\*x$. The performance of this approach is clearly affected by the choice of the observation that is used to construct the initial posterior distribution. That is, the selection of the datum $x_1$ will impact the length of the resulting code, with some data $x_1$ resulting in shorter codes than others.
\paragraph{Approximate objective Bayesian code}
The approximate objective Bayesian code is computed using the same principle as the objective Bayesian code. We first compute the Jeffreys posterior distribution using the datum $x_1$ and then use this as a prior distribution for data $\*x_2^n$. The difference arises when computing the codelength, where an asymptotic formula under suitable regularity conditions \citep{balasubramanian1997statistical,clarke1990information} is used in place of the exact negative logarithm of the marginal likelihood (\ref{eq:marginal}): 
\begin{equation*}
	%\label{eq:marginal.asym}
	-\ln m(\*x) = -\ln{f(\*x|\theta)} + \frac{k}{2}\ln\frac{n}{2\pi} -\ln \pi(\theta) + \frac{1}{2}\ln |F(\theta)| \, + o(1). 
\end{equation*}
The difference between the objective Bayesian and the approximate objective Bayesian codelengths tends to $0$ as the sample size $n$ tends to infinity. The codelengths for the Poisson and geometric models under the approximate objective Bayesian code are
\begin{align*}
I_{P}(\*x_2^n|x_1) &= l_{P}(\*x_2^n|\hat{\mu}_2^n) + \frac{1}{2}\ln\frac{n}{2\pi} + \hat{\mu}_2^n - x_1\ln\hat{\mu}_2^n + \ln\Gamma\left(x_1+\tfrac{1}{2}\right) \,, \\
I_{G}(\*x_2^n|x_1) &= l_{G}(\*x_2^n|\hat{\mu}_2^n) + \frac{1}{2}\ln\frac{n}{2\pi} + x_1\ln\left(1+\frac{1}{\hat{\mu}_2^n}\right) + \frac{1}{2}\ln\hat{\mu}_2^n - \ln\left(x_1 + \tfrac{1}{2}\right) \,,
\end{align*}
where $\*x_2^n = (x_2,\dots,x_n)$ is the data without the first observation and $\hat{\mu}_2^n$ is the MLE computed using data $\*x_2^n$. As with the exact objective Bayesian code, the performance of the approximate objective Bayesian code depends on the choice of the datum $x_1$.
\section{Minimum Message Length}
\label{sec:mml}
\subsection{Introduction to MML}
Minimum message length (MML) model selection was introduced by C. S. Wallace and D. M. Boulton~\citep{wallace2005statistical,wallace1968information}. Like MDL, MML is an inductive inference method based on data compression. Suppose we are given some data $\*x$ that we would like to send to an imaginary receiver by encoding it into a message (e.g., a binary string). This MML message consists of two parts: (1)~a description of a model $\theta \in \Theta^* \subset \Theta$, and (2)~a description of the data using the model $f(\*x | \theta)$ specified in the first message component. 
%\begin{enumerate}
%	\item The transmitter first selects a model characterized by the parameter $\theta \in \Theta^* \subset \Theta$
%	\item Then the transmitter would send the data using that model 
%\end{enumerate}
The set $\Theta$ denotes the parameter space for the statistical model $f(\cdot)$, and $\Theta^*$ is a countable subset of the parameter space containing all possible MML estimates $\hat{\theta}$ that can be used to transmit the data. The coding scheme (i.e., the set $\Theta^*$ of MML estimates and corresponding codewords) is agreed upon by both the sender and receiver before any data is seen. 

In MML terminology, the first part of the message, which encodes the model structure and the model parameters, is called the \emph{assertion}. The second part of the message is called the \emph{detail} and encodes the observed data $\*x$ using the model specified in the assertion. In order to create codewords for the set $\Theta^*$, MML requires a prior probability distribution $\pi(\cdot)$ over $\Theta$ and is therefore a Bayesian procedure. The total message length of the data $\*x$ and a model $\theta \in \Theta^*$ is
\begin{equation*}
	I(\*x,\theta) = \underbrace{I(\theta)}_{\text{assertion}} + \;\; \underbrace{I(\*x|\theta)}_{\text{detail}} \,.
\end{equation*} 
The length of the assertion is a measure of the model complexity, while the length of the detail is measure of the goodness-of-fit of the model to the data. MML seeks the model that minimizes this tradeoff between model complexity and model capability, i.e., 
\begin{equation*}
	\hat{\theta}_{\text{MML}}(\*x) = \argmin_{\theta \in \Theta^*} \{I(\theta) + I(\*x|\theta)\} \,.
\end{equation*}
The key step in MML inference is the construction of the countable set $\Theta^*$ and associated codewords for members of this set. In the strict minimum message length (SMML) approach~\citep{wallace1975invariant}, the set $\Theta^*$ is obtained by minimizing the expected codelength of data under the assumption that the data comes from the marginal distribution (\ref{eq:marginal}). Exact solutions to this optimization problem are in general NP-hard except for the case of very simple problems~\citep{farr2002complexity}. 

To address this problem, several computationally tractable SMML codelength approximations have been developed~\citep{wallace2005statistical,dowe2008foreword,schmidt2011new}, the most popular approximation being the MML87 approximation~\citep{wallace1987estimation}. The MML87 message length for a model parameterised by $\theta \in \mathbb{R}^k$ is  
\begin{equation}
\label{eq:mml87_general}
	I(\*x, \theta) = \underbrace{-\ln \pi(\theta) + \frac{1}{2}\ln\abs{F(\theta)} + \frac{k}{2} \ln \kappa_k}_{\text{assertion}} + \underbrace{\frac{k}{2} -\ln f(\*x|\theta)}_{\text{detail}} \,,
\end{equation}
where $\pi(\theta)$ is a prior density for $\theta$, $F(\theta)$ is the Fisher information for $n$ data points, ${f(\*x|\theta)}$ is the sampling density of the model, and $\kappa_k$ is is the normalized second moment of an optimal quantizing lattice and can be approximated \citep{wallace2005statistical} by
\begin{equation}
\label{eq:kappa.approx}
\frac{k}{2}(\ln\kappa_k+1) \approx -\frac{k}{2}\ln(2\pi) + \frac{1}{2}\ln(k\pi) + \psi(1) \,,
\end{equation}
where $\psi(\cdot)$ is the digamma function. The MML estimator is defined as the $\hat{\theta}$ that minimizes (\ref{eq:mml87_general}). The message length, evaluated at the minimum $\hat{\theta}$ denotes the codelength of the shortest two-part message that can be used to encode both the data and the model given the prior distribution $\pi(\theta)$. For models with one free parameter, such as the Poisson and geometric distributions, the MML87 approximation (\ref{eq:mml87_general}) simplifies to
\begin{equation}
	\label{eq:mml87}
	I(\*x,\theta) = \underbrace{-\ln{\pi(\theta)} + \frac{1}{2}\ln |F(\theta)| - \frac{1}{2}\ln{12}}_{\text{assertion}} + \underbrace{\frac{1}{2} -\ln{f(\*x|\theta)}}_{\text{detail}} \, ,
\end{equation}
where $\kappa_1 = 1/12$ is the optimal quantization constant in one dimension.

The key reason that MML87 is computationally tractable is that it avoids explicitly constructing the quantized parameter space $\Theta^*$. Instead, for a given $\theta^\prime \in \Theta^*$, MML87 finds the (approximate) subset of parameters in $\theta \in \Theta$ that are closer in some sense to $\theta^\prime$ than to any other member of $\Theta^*$. This subset of parameters is called the \emph{uncertainty region} and its volume $w(\theta)$ is given by 
\begin{equation*}
	w(\theta) = \left( |F(\theta)| \, \kappa_k^k \right)^{-\frac{1}{2}},
\end{equation*}
and the length of the assertion in MML87 is therefore 
\begin{equation*}
	I(\theta) = -\ln \pi(\theta) w(\theta).
\end{equation*}
The size of the uncertainty region $w(\theta)$ depends on the variation of the likelihood function around $\theta$. If a small change in $\theta$ results in a large increase in the negative log-likelihood of the data, the uncertainty region will be small. Conversely, if the negative log-likelihood is insensitive to small changes in $\theta$, the uncertainty region will be large. From Shannon's theory of information (\ref{eq:Shannon}) we see that the MML87 approximation assigns longer codewords, and therefore greater complexity, to models that need to be more precisely specified, i.e., those with small uncertainty regions. Given two candidate models that fit the observed data equally well, MML advocates choosing the model with a larger uncertainty region. In this case, a model with a small uncertainty region fits the observed data well but is less likely to generalize to unseen data than a model with a large uncertainty region.

%If all the members of $\Theta^*$ were known, we could divide $\Theta$ into a set of regions where each region is associated with one member of $\Theta^*$. The volume of these regions determines the codewords assigned to each $

\paragraph{Example}
Consider an experiment with $n$ independent Bernoulli trials $\*x=(x_1,\dots,x_n)$ and let $p$ denote the probability of success in each trial. Suppose we model $p$ using a uniform prior $\pi(p) = 1$. Using (\ref{eq:mml87}), the MML87 message length of the data $\*x$ and the model $p$ is 
\begin{equation}
	\label{eq:MMLBinomial}
	I(\*x,p) = -n_1 \ln p - (n-n_1) \ln(1-p) + \frac{1}{2} \ln \left(\frac{n}{p(1-p)}\right) - \frac{1}{2}\ln{12} + \frac{1}{2},
\end{equation}
where $n_1$ is the number of successes. By minimizing (\ref{eq:MMLBinomial}) with respect to $p$, we obtain the MML87 estimate
\begin{equation*}
	\hat{p}_{\tiny{\text{MML}}}(\*x) = \frac{n_1 + \frac{1}{2}}{n + 1} \,.
\end{equation*}
In comparison to the maximum likelihood estimate $n_1 / n$, the MML87 estimate $\hat{p}_{\tiny{\text{MML}}}$ is always closer to $p = 1/2$ and can never take on the values $p=0$ or $p=1$. This is because the likelihood of models near the boundary is sensitive to small changes in $p$. MML assigns greater assertion lengths, and therefore greater complexity, to these models in comparison to models near $p=1/2$. To see this, note that the volume of the uncertainty region for the Bernoulli model is
\begin{equation*}
	w(p) = \sqrt{\frac{12p(1-p)}{n}} \, ,
\end{equation*}
which decreases as the sample size $n$ grows or as the success probability $p$ gets closer to the parameter space boundary.

\begin{figure}[h]
\centering
\includegraphics[width=\textwidth]{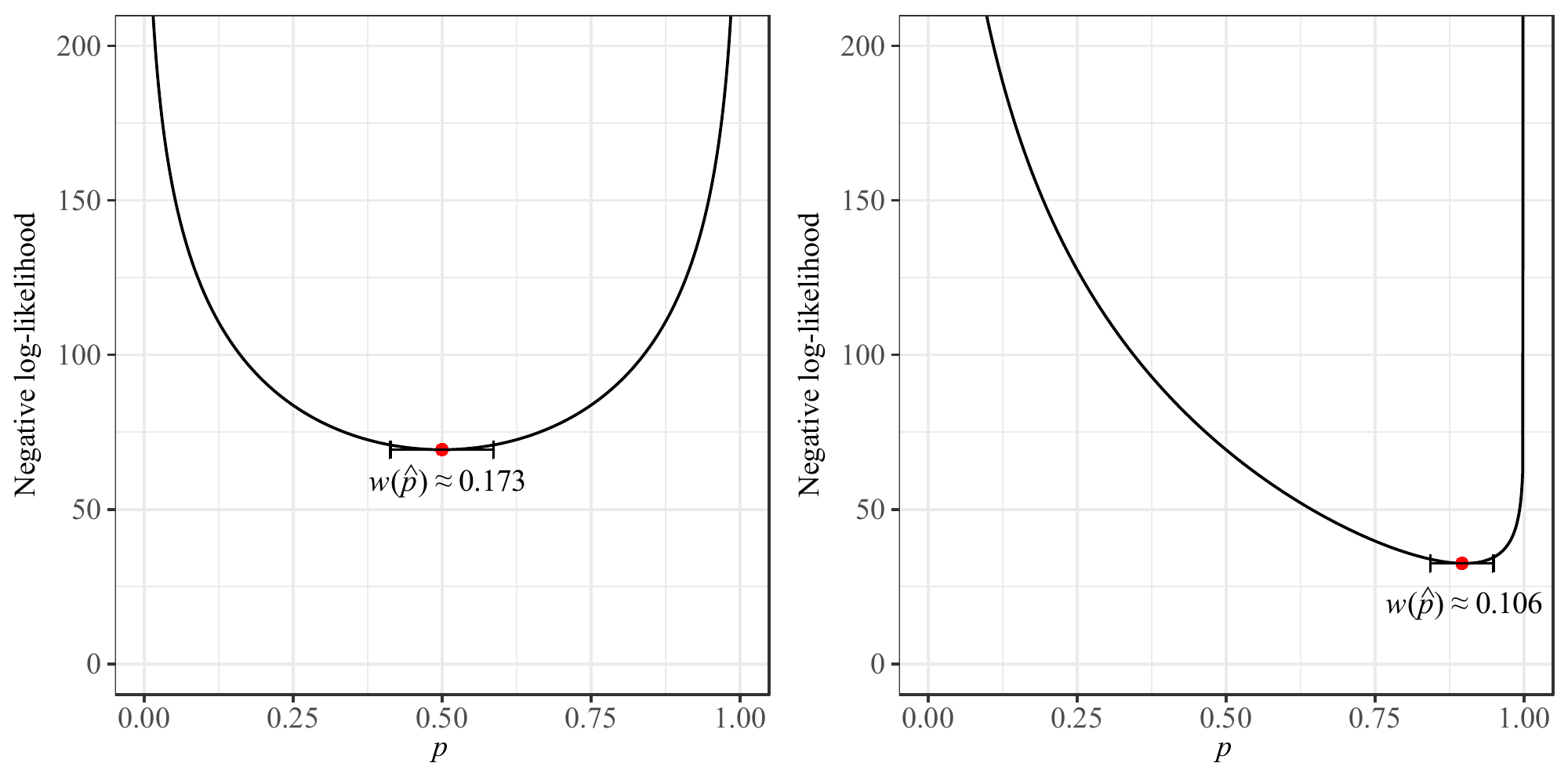}
\caption{Plot of negative log-likelihood against $p$ in $n=100$ independent Bernoulli trials. The observed counts of success are $n_1=50$ (left) and $n_1=90$ (right). The dots are the MML estimates and the error bars represent the corresponding uncertainty regions.}
\label{fig:mml_bin}
\end{figure}

Figure \ref{fig:mml_bin} shows the plots of the negative log-likelihood $-\ln f(n_1|p)$ against the success probability $p \in (0,1)$ for two independent Bernoulli data sets, each of size $n=100$. The observed counts of success were $n_1 = 50$ (left panel) and $n_1 = 90$ (right panel). The MML87 estimates for these data sets are represented by the red dots, and the error bars show the width of the uncertainty region associated with each estimate. We observe that in the case when $n_1=50$ the MML87 estimate is $\hat{p}_{\tiny{\text{MML}}} = 0.5$ and the uncertainty region is large. In other words, we do not need to encode $\hat{p}_{\tiny{\text{MML}}}$ to a high precision, since small changes to $\hat{p}_{\tiny{\text{MML}}}$ will not greatly affect the likelihood. In contrast, when $n_1=90$ the MML87 estimate is $\hat{p}_{\tiny{\text{MML}}} \approx 0.9$, the uncertainty region is smaller and $\hat{p}_{\tiny{\text{MML}}}$ should be encoded to higher precision. Therefore, MML87 assigns greater complexity to estimates close to the boundary of the parameter space.
% small change in $p$ leads to a larger change of the likelihood, especially when the true $p$ is near $1$. Therefore, the uncertainty region is shorter, and we need to encode $p$ with a higher precision. 

\paragraph{Invariance}
An attractive property of MML is that it is model invariant. Suppose the model class, which is originally parameterized by $\theta$, is given a new parameterization $\phi=g(\theta)$, where $g(\cdot)$ is an invertible function. If $\hat{\theta}$ is the MML estimate which minimizes the MML87 message length formula (\ref{eq:mml87_general}), then $\hat{\phi} = g(\hat{\theta})$ is the MML estimate of $\phi$ in this new parameterization. Note that the MLE is also invariant to re-parametrization while the standard Bayesian posterior mode (MAP) and posterior mean estimators are, in general, not invariant.

\paragraph{MML87 and the Jeffreys prior}
Suppose we use the Jeffreys prior distribution for $\theta$ given by
\begin{equation*}
	\pi(\theta) = \frac{\sqrt{|F_1(\theta)|}}{\int_{\Theta} \sqrt{|F_1(\theta^\prime)|} d\theta^\prime}  \,,
\end{equation*}
where $|F_1(\theta)|$ is the determinant of the Fisher information matrix for a single data point. Using this prior in conjunction with the approximation (\ref{eq:kappa.approx}), the MML87 message length formula (\ref{eq:mml87_general}) is
\begin{equation*}
	I(\*x,\theta) = 
	-\ln f(\*x|\theta) + \ln\int_{\Theta} \sqrt{|F_1(\theta^\prime)|} d\theta^\prime + \frac{k}{2}\ln \frac{n}{2\pi} + \frac{1}{2}\ln(k\pi) + \psi(1) \, .
\end{equation*}
%
%\begin{equation*}
%	I(\*x, \theta) = - \ln f(\*x|\theta) + \ln\int_{\Theta} \sqrt{|F_1(\theta)|} + \frac{k}{2}\ln n + \frac{k}{2} \ln\kappa_k + \frac{k}{2} \,,
%\end{equation*}
%
Under the Jeffreys prior, the MML87 estimate is equivalent to the MLE and the MML codelength is similar to the NML codelength (\ref{eq:nml.code}) with the parametric complexity approximation (\ref{eq:pc.approx}). The MML codelength is slightly longer than the NML codelength, and the difference is  
\begin{equation*}
	I(\*x,\hat{\theta}_\textrm{\tiny{MML}} ) + \ln p_\textrm{\tiny{NML}}(\*x) = O( \ln k ) \,.
	%\frac{1}{2}\ln(k\pi) + \psi(1) + o(1)\,.
\end{equation*}
The MML codelength is necessarily longer than the one-part NML codelength as MML is based on two-part codes that always assert a fully specified model (i.e., model class and parameter estimates). This extra codelength allows MML to perform both parameter estimation and model selection within the same information-theoretic framework which is not possible using one-part codes.

\paragraph{Applications}
MML principle has been applied widely across different areas in statistics and computer science. Some examples of successful applications include linear regression~\citep{schmidt2009mml}, decision trees~\citep{wallace1993coding}, causal models~\citep{wallace1996causal}, time series~\citep{fitzgibbon2004minimum, schmidt2013minimum}, neural networks~\citep{makalic2004rbf} and mixture modeling~\citep{wallace2000mml}. A more extensive list of MML applications is available in~\cite{wallace2005statistical}.

%In this paper, the parameter $\theta$ will be corresponded to $\lambda$ in the Poisson and $p$ in the geometric model. 
%\begin{exmp}
%	Binomial distribution. Consider an experiment of tossing a coin $n$ times and the observed data is denoted by $\*x=\{x_1,\dots,x_n\}$. Assume 
%\end{exmp}

\subsection{MDL and MML: similarities and differences}
The MML and MDL principles have a number of important characteristics in common. In particular, both principles are based on the insight that structure can be learned by compressing data. In both approaches, the best hypothesis is the model that most compresses the data (i.e., leads to the shortest codelength of the data and the model). However, there exist some important differences between the two model selection principles. First, MDL and MML infer different types of models. MDL aims to infer the best model class but does not nominate a fully specified model (i.e., a particular member of that model class) and implicitly endorses the maximum likelihood estimator. In contrast, MML is based on two-part codes and always nominates a fully specified model (i.e., both the model class and parameter estimates). In many problems MML estimators have demonstrated improved empirical performance when compared to maximum likelihood and standard Bayesian estimators. Second, MML constructs codes that minimize the expected codelength where the expectation is taken with respect to the marginal distribution of the data. In contrast, MDL constructs codes that minimize the worst-case codelength relative to the ideal code (i.e., the regret (\ref{eq:regreg})). Unlike MML, the MDL principle is strictly non-Bayesian and attempts to avoid any use of subjective prior information when constructing codes. 

However, when the NML parametric complexity term is infinite, researchers resort to restricting the range of integration such that the resultant criterion is finite, for example, see Section~\ref{sec:mdl.pois.geom}. The choice of this restricted parameter set is essentially equivalent to choosing a prior distribution. Restricting the parameters to a particular region has an effect on the NML codelength that is difficult to interpret, particularly in models with more than one free parameter. In contrast, the effect of the prior distribution on the MML codelength is significantly more transparent. A further advantage of the explicitly Bayesian nature of MML is that researchers can draw on the extensive body of Bayesian literature when specifying appropriate priors.

%For models with more than one parameter restricting the parameter space is even more difficult.
%The usage of prior information is avoid in the MDL framework. 
%No informative priors are put on the model classes, they all have the same assertion length. 
%Stochastic Complexity

Both MDL and MML are capable inference methods and have been shown to have excellent performance in many applications. Although they are philosophically different, both approaches often obtain similar results in practice. Further discussion of the MML and MDL similarities and differences can be found in~\cite{baxter1994mdl} and \cite[pp. 413 -- 415]{wallace2005statistical}.
%
%However, there exist some important differences between the two model selection principles. One difference between MML and MDL is that, because MML is a Bayesian principle, a prior distribution for all parameters is always required. A second difference between MML and MDL is that MML always encodes a two-part message and unlike MDL, which is based on one-part codes, the MML principle can be used to perform both parameter estimation and model selection. MML is parameterisation-invariant, and such makes the same inferences irrespective of which parameterisation of a model is used. 
%For example, although we used a different parameterisation from~\cite{de2006empirical} for the geometric model, our model selection procedure will yield the same results.
%
%
% Compared with the restricted ANML codes (\ref{eq:rnml.pois}) and (\ref{eq:rnml.geom}), in which we have to set a parameter region, it can be argued that the use of priors in MML is more direct and easier to interpret.
%
\subsection{Prior Distributions for Poisson and geometric models}
\label{ssec:priors}
%
%%!! ADD TEXT: SOME INTRODUCTORY TEXT IS NEEDED HERE. SAY MML IS BAYESIAN, WE NEED PRIORS FOR THETA. THEN SAY THAT WE WILL LOOK AT 3 DIFFERENT TYPES OF PRIORS, ETC.
Since MML is a Bayesian principle, prior distributions for all model parameters are required. In this section, we will examine three choices of prior distributions for the Poisson and geometric models. We start by presenting the conjugate priors (Section \ref{sssec:expbeta}), which are a common choice in Bayesian statistics due to their mathematical convenience. In Section \ref{sssec:match1st} we present a method for calibrating these conjugate prior distributions for the geometric and Poisson by matching moments. Lastly, we examine the use of the half-Cauchy prior (Section \ref{sssec:cauchy}), which in contrast to the conjugate priors is heavy-tailed and free of any user chosen hyperparameters. 
%
%
%We then introduce the mean matching priors (Section \ref{sssec:match1st}), which is similar to the conjugate priors, with the constraint that the mean of the two models are match. 
%
\subsubsection{Prior \RN{1}: Conjugate priors}
\label{sssec:expbeta}
We select an exponential prior for the Poisson model and a beta prior for the geometric model as these distributions are conjugate for the Poisson and geometric models, respectively. Suppose $\lambda \sim \textrm{Exp}(1/A)$ for some $A>0$ and $p \sim \textrm{Beta}(\alpha,\beta)$ for some $\alpha,\beta >0$, where $p$ is the success probability for the geometric distribution. The prior densities for $\lambda$ and $p$ are
%Using a conjugate prior is mathematically convenient. 
%
\begin{align}
% h_P(\lambda) &= (1/A)\exp(-\lambda/A) \,, \\ 
% h_G(p) &= \frac{p^{\alpha-1}(1-p)^{\beta-1}}{\text{B}(\alpha,\beta)} \,, \\
	\pi_P(\lambda) &= \frac{1}{A}\me^{-\tfrac{\lambda}{A}} \,, \label{eqn:poiss:prior} \\
	\pi_G(p) &= \frac{1}{\text{B}(\alpha,\beta)} p^{\alpha-1}(1-p)^{\beta-1} \,. \label{eqn:geo:prior}
\end{align}
where $B(\cdot)$ is the beta function. The values of $\lambda$ and $p$ that minimize (\ref{eq:mml87}) using these prior distributions are
\begin{align}
	\label{eq:est.pois.mmlexpbeta}
	\hat{\lambda}_\textrm{\tiny{MML}}(\*x) &= \frac{s + 1/2}{n+1/A} \,, \\
	\label{eq:est.geom.mmlexpbeta}
	\hat{p}_\textrm{\tiny{MML}}(\*x) &= \frac{n+\alpha}{n+\alpha+\beta+s- 3/2} \,.
\end{align}
The minimum codelength is then obtained by plugging $\hat{\theta}_\textrm{\tiny{MML}}$ into (\ref{eq:mml87}), where $\hat{\theta}_\textrm{\tiny{MML}} = \hat{\lambda}_\textrm{\tiny{MML}}$ for the Poisson and $\hat{\theta}_\textrm{\tiny{MML}} = \hat{p}_\textrm{\tiny{MML}}$ for the geometric model. 

The MML estimates for both the Poisson and geometric models are equivalent to MLEs based on augmented data. In the case of the Poisson model, the MML estimator (\ref{eq:est.geom.mmlexpbeta}) is equivalent to the MLE of a dataset augmented with $1/A$ additional fake data points whose sum is equal to $1/2$. In the case of the geometric model, the MML estimator (\ref{eq:est.geom.mmlexpbeta}) is equivalent to the MLE of a dataset augmented with $\alpha$ additional fake data points whose sum is equal to $(\beta - 3/2)$.  In the limit, for large sample sizes the MML estimators for both Poisson and geometric models converge to their respective MLEs.
\subsubsection{Prior \RN{2}: Calibrated conjugate priors}
\label{sssec:match1st}
The conjugate priors we used in Section \ref{sssec:expbeta} have a potential problem. Assume we are not given any information regarding which model is better for explaining the observed data. If we use fixed values for the hyperparameters $A$, $\alpha$ and $\beta$, our model selection procedure may favor one model over another due to this arbitrary choice of priors. As an alternative, we propose to calibrate the conjugate priors distributions (\ref{eqn:poiss:prior}) and (\ref{eqn:geo:prior}) so that the prior expected means of the two models are equal. The mean of the Poisson model is $\lambda$ and the mean of the geometric model is $(1-p)/p$. If $\lambda$ follows the exponential prior (\ref{eqn:poiss:prior}), then the prior expected mean of the Poisson model is $\EX_\lambda(\lambda) = A$. Given a value for the fixed hyperparameter $A$, we can calibrate the conjugate prior for the geometric parameter $p$ (\ref{eqn:geo:prior}) such that the prior expected mean of the geometric model is equal to $A$, i.e., find the values of $\alpha$ and $\beta$ such that $\EX_p \left[(1-p)/p\right] = A$. This expectation is given by
\begin{align}
\label{eq:geom.mean.mean}
\EX_p \left(\frac{1-p}{p}\right) &= \frac{\beta}{\alpha-1} \,,
\end{align}
and the values of $\alpha$ and $\beta$ that calibrate the two conjugate priors are
\begin{equation*}
	\alpha = \beta = \frac{A}{A-1} \,.
\end{equation*}
The prior distributions we use for the parameters $\lambda$ and $p$ are then
\begin{align}
	\lambda &\sim \textrm{Exp}\left(\frac{1}{A}\right) \,, \\
	p &\sim \textrm{Beta} \left( \frac{A}{A-1},\frac{A}{A-1} \right) \,. \label{eq:geom:calibrated}
\end{align}

%Two observations can be made here. First, 
From (\ref{eq:geom.mean.mean}), putting a uniform prior on $p$ by setting $\alpha=\beta=1$ results in the prior expected mean of the geometric model being infinite. Conversely, the prior expected mean of the Poisson model is $\EX_\lambda(\lambda) = A$ which is always finite. From (\ref{eq:geom:calibrated}), we see that the larger the value of $A$ used in the exponential prior, the more similar the MML estimator with prior \RN{2} behaves to an MML estimator with a uniform prior on $p$, since $A/(A-1) \to 1$ as $A \to \infty$. 

% so $p \approxd \textrm{Unif}(0,1)$ when $A$ is large.

%This is expected since coding the data under the Poisson model is inefficient compared with the geometric model as the sufficient statistic $s = \sum_{i=1}^n x_i$ grows and moves far away from $A$. Second, we expect that the larger the value of $A$ in the exponential prior, the more similar the MML estimator using prior \RN{2} behaves to an MML estimator with a uniform prior on $p$, since $A/(A-1) \to 1$ as $A \to \infty$, so $p \approxd \textrm{Unif}(0,1)$ when $A$ is large.%%!! GIVE MORE REASON WHY THIS IS THE CASE. YOU COULD SAY HERE THAT lim A->inf of A/(A-1) -> 1 which means p ~ Unif(0,1).
%
%
%
\subsubsection{Prior \RN{3}: Half-Cauchy prior}
\label{sssec:cauchy}
The prior distributions proposed in Section~(\ref{sssec:expbeta}) and (\ref{sssec:match1st}) require the selection of the hyperparameter $A$ which controls the {\em a priori} expected value of $\lambda$. This is not easy in practice if we do not have actual prior knowledge of the data generating process. Instead, we propose to use a prior distribution that is free of user-chosen hyperparameters. The idea is to use the half-Cauchy distribution~\citep{polson2012half} as the prior distribution on the standard deviation of both models
\[
	\sqrt{\lambda} \sim \text{C}^+(0,1) \,, \; \; \;\sqrt{\frac{1-p}{p^2}} \sim \text{C}^+(0,1) \,. 
\]
The half-Cauchy prior is a recommended default choice for scale parameters and has heavy polynomial tails~\citep{gelman2006prior}. Given the above priors for the standard deviation, the corresponding prior densities for $\lambda$ and $p$ are
\begin{align}
	\label{eq:half_cauchy_lambda}
	\pi_\lambda(\lambda) &= \frac{1}{\pi \sqrt{\lambda}(1+\lambda)} \,, \\
	\label{eq:half_cauchy_p}
	\pi_p(p) &= \frac{2-p}{\pi \sqrt{1-p} (p^2-p+1)} \,.
\end{align}

Following the above approach, it is also possible to directly model the (square root) of the mean of the geometric distribution, $\mu$, using the half-Cauchy prior. This can easily be achieved by using the conjugate beta prior for the parameter $p$ discussed in Section~\ref{sssec:match1st} and setting the hyperparameters $\alpha = \beta = 1/2$ which yields
\begin{equation}
	\sqrt{\mu} = \sqrt{\frac{1-p}{p}} \sim \text{C}^+(0,1) \,. 
\end{equation}
The mean and variance of a Poisson distribution are both equal to $\lambda$ and therefore using a half-Cauchy prior on the square root of the mean is equivalent to using a half-Cauchy prior on the standard deviation (\ref{eq:half_cauchy_lambda}). It is not clear whether it is preferable to match the Poisson and geometric distributions using prior distributions over their standard deviations or their means. One advantage of the latter approach is that the resultant MML estimate for the geometric distribution retains the simple analytical solution given by (\ref{eq:est.geom.mmlexpbeta}).

In the case of the Poisson distribution, the MML estimator for $\lambda$ using the prior (\ref{eq:half_cauchy_lambda}) is
\begin{equation}
	\label{eq:est.pois.mmlcauchy}
	\hat{\lambda}_\textrm{\tiny{MML}}(\*x) = \left(\frac{1}{2n}\right) \left( \sqrt{s^2+2s(n-1)+(n+1)^2}+s-n-1 \right) \,,
\end{equation}
where $s=\sum_{i=1}^n x_i$. In case of the geometric distribution, the MML estimator using (\ref{eq:half_cauchy_p}) is a solution of the quartic polynomial
\begin{equation}
	\label{eq:geom.quartic}
	q_G(p) = -(s+n)p^4 + (3s+4n-1)p^3 - (3s+6n+1)p^2 + (2s+5n+4)p - 2n -2 \,,
\end{equation}
and may be obtained numerically. 
\subsection{Discussion}
\label{ssec:remark.prior}
\paragraph{MML codelengths}
\label{ssec:mmlcodelengths}
MML inference with prior distribution \RN{1} (Section \ref{sssec:expbeta}) and \RN{2} (Section \ref{sssec:match1st}) requires specification of the fixed hyperparameter $A$ which has two disadvantages. First, specifying a value for the hyperparameter $A$, before any data is observed, is difficult in practice. Second, MML with either prior distributions \RN{1} or \RN{2} exhibits strong model selection bias towards the geometric model when the hyperparameter $A$ differs significantly from the sample mean of the data. 

\begin{figure}[h]
\centering
\begin{subfigure}{.5\textwidth}
  \centering
  \includegraphics[width=1\linewidth]{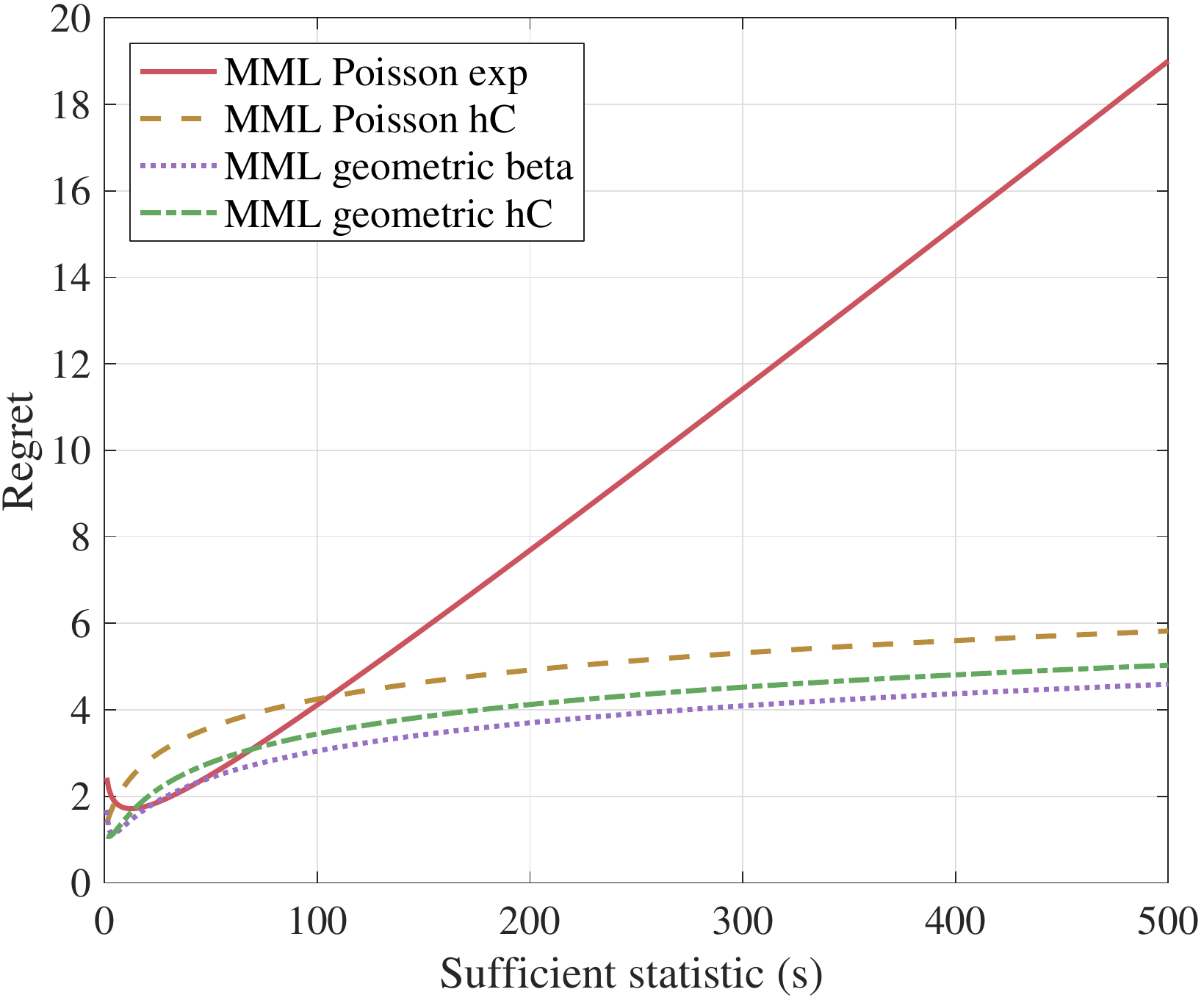}
  \caption[short]{MML regret for the Poisson and geometric models}
  \label{fig:mml_regret}
\end{subfigure}%
\begin{subfigure}{.5\textwidth}
  \centering
  \includegraphics[width=1\linewidth]{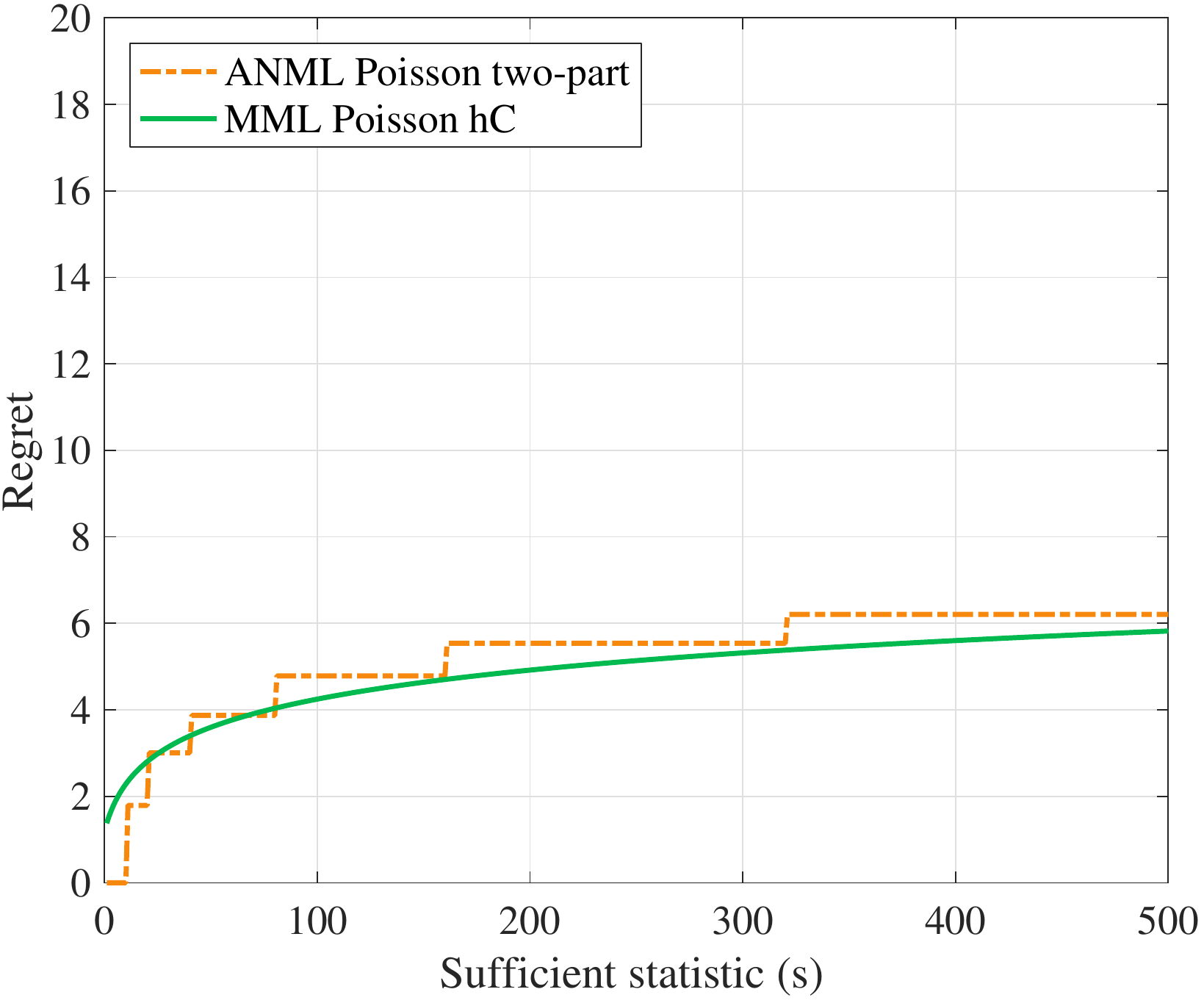}
  \caption[short]{MML and ANML two-part regret in the Poisson model}
  \label{fig:nml_regret}
\end{subfigure}
\caption{Plot of the MML and two-part ANML coding regret against the sufficient statistic in the Poisson and geometric models. Priors for the Poisson model are an exponential prior with parameter $A=5$ and a half-Cauchy (hC) prior on the standard deviation. Priors for the geometric model are a beta prior with parameter $\alpha=1$, $\beta=1$ and a half-Cauchy prior on the standard deviation.}
\label{fig:regret_plot}
\end{figure}

To understand this behavior, we investigate the coding regret of the MML criterion under the proposed priors. Figure~\ref{fig:mml_regret} shows a plot of the regret (\ref{eq:regreg}) of the MML codelengths using priors \RN{1} and \RN{3} against the sufficient statistic $s = \sum_{i=1}^n x_i$. The regret of the MML codelength using the exponential prior distribution grows at a much faster rate (linear in $s$) than the regret of the MML codelength using the half-Cauchy prior \RN{3} (logarithmic in $s$). In contrast, for the geometric model, both the beta prior distribution, irrespective of $\alpha$ and $\beta$, and the half-Cauchy prior approach yield a logarithmic rate of regret growth as $s \to \infty$. Coding the data under the Poisson model with an exponential prior becomes increasingly inefficient for all $s/n$ much greater than $A$. Therefore, if the data is generated by a model whose mean is far away from the fixed hyperparameter $A$, the MML code with the proposed conjugate priors \RN{1} or \RN{2} will tend to favor the geometric model over the Poisson model. In contrast, a standard likelihood ratio test would generally favor the Poisson model over the geometric model as the Poisson has higher parametric complexity. This emphasizes the importance of selecting appropriate prior distributions when using Bayesian techniques such as MML and the potential benefits of using heavy-tailed priors.
\paragraph{Comparison of NML and MML}
As both MML and MDL criteria are model selection techniques based on data compression we can analyze their behavior in terms of coding regret (\ref{eq:regreg}). As previously discussed, the MML criterion with the half-Cauchy prior for both the Poisson and geometric models attains a logarithmic rate of increase in regret as $s \to \infty$. In contrast, the restricted ANML criterion is not defined for data where the sufficient statistic $s$ is greater than $n \mu^*$ and therefore has infinite regret. The two-part restricted ANML code circumvents this problem by estimating $\mu^*$ from the observed data and thus attains finite regret for all data.

Figure~\ref{fig:nml_regret} shows a plot of the MML and two-part restricted ANML coding regret against the sufficient statistic $s$ for the Poisson model. Both the MML and NML criteria attain a logarithmic increase in the coding regret as $s \to \infty$. However, in the case of the MML half-Cauchy, the regret increase is a smooth function of $s$ while the use of the $\log^*$ code in the NML criterion results in jump discontinuities. In the case of the geometric distribution, the MML criteria based on priors \RN{1}--\RN{3} and the two-part restricted ANML code attain a logarithmic increase in the coding regret as $s \to \infty$. In light of these observations, we expect the two-part restricted ANML criterion and the MML criterion with the half-Cauchy prior to perform similarly when used to discriminate between Poisson and geometric models.
%expected to behave similarly when used for model selection
%
%the two-part restricted ANML and the half-Cauchy 
%
%
%In comparison to the NML with the geometric, A similar result holds in the case of the geometric distribution with both conjugate or half-Cauchy prior distributions.
%
%We see that the regret of MML with an exponential prior diverges as $s$ increases, compared with a much slower divergence rate for MML with a Cauchy prior. Furthermore, we see that the rate of the regret for the MML and two-part ANML methods is similar with MML Cauchy having a smooth curve as opposed to the jump discontinuities observed in the two-part ANML approach.
%
%
%Compared with the restricted ANML codes (\ref{eq:rnml.pois}) and (\ref{eq:rnml.geom}), in which we have to set a parameter region, it can be argued that the use of priors in MML is more direct and easier to interpret.
%

\paragraph{MML parameter estimators}
For the Poisson model with an exponential prior, the MML parameter estimate for $\lambda$ is known to be asymptotically biased as $s \to \infty$. The bias relative to the MLE is given by
\begin{equation*}
	\lim_{s \to \infty} \left\{ \frac{ \hat{\lambda}_\textrm{\tiny{MML}} } {\hat{\lambda} } \right\} = \frac{n}{n + 1/A} \,,
\end{equation*}
which is always less than one for all finite values of $A$ and $n$. As discussed by~\cite{CarvalhoPolson10}, an estimator with a relative bias that does not tend to unity as $s \to \infty$ is undesirable since the absolute bias of the estimator increases proportionally with increasing $s$. In contrast, the MML estimate of the Poisson parameter $\lambda$ using the half-Cauchy prior (\ref{eq:est.pois.mmlcauchy}) has an asymptotic relative bias of one since
\begin{equation}
	\label{eqn:poisson:relativebias}
	\lim_{s\to\infty} \left\{ \frac{\hat{\lambda}_\textrm{\tiny{MML}}}{\hat{\lambda}} \right\} = \lim_{s\to\infty}\left\{ \frac{1}{2} \left( \sqrt{1+\frac{1}{s}(3n-1)}+1-\frac{1}{s}(n+1) \right) \right\} = 1 \,.
\end{equation}
The absolute bias of the MML estimate using the half-Cauchy prior is of order $O(1)$. As most loss functions used to assess the quality of estimating the Poisson rate parameter are normalized by $\lambda$, this bias disappears for large values of $\lambda$ since ${\rm E}(\hat{\lambda} - \hat{\lambda}_\textrm{\tiny{MML}}) = o(\lambda)$.

%this is bad because since as s grows the bias gets worse for any finite n

%we first show the bias of the MML estimator as s goes to infinity

%eqn: lim lambda mml / lambda ml

%In contrast, the MML estimate with the half-cauchy prior has relative bias of zero/one compared to MML. it may have constant bias in s, but this is fine
%most error measures of estimating lambd are normalised by lambda

%geometric model

%From (\ref{eq:est.pois.mmlcauchy}), we expect that for a fixed sample size $n$, as $s \to \infty$, this MML estimator has the same behavior as the MLE since 
%
%\begin{equation*}
%	\lim_{s\to\infty} \left\{ \frac{\hat{\lambda}_\textrm{\tiny{MML}}}{\hat{\lambda}_\textrm{\tiny{MLE}}} \right\} = \lim_{s\to\infty}\left\{ \frac{1}{2} \left( \sqrt{1+\frac{1}{s}(3n-1)}+1-\frac{1}{s}(n+1) \right) \right\} = 1 \,.
%\end{equation*}
%

In the case of the geometric distribution, the MML estimator with the half-Cauchy prior on the standard deviation is defined in terms of the solution to the quartic polynomial (\ref{eq:geom.quartic}). For large values of $s$, the MML estimate of $p$ will tend to zero since the only permissible root of the polynomial
\begin{equation*}
	%\label{eq:geom.quartic.sinf}
	\lim_{s \to \infty} \left\{ \frac{q_G(p)}{s} \right\} = p^4+3p^3-3p^2+2p \,,
\end{equation*}
is $p=0$. Furthermore, for large $s$, the first three terms in (\ref{eq:geom.quartic}) are negligible compared with the last three terms, which means that the MML estimator can be approximated by
\begin{equation*}
	\hat{p}_\textrm{\tiny{MML}}(\*x) \approx \frac{n+1}{(5/2)n + s + 2} \,.
\end{equation*}

The bias relative to the MLE for both the MML estimator with the beta prior on the probability $p$, and the MML estimator with a half-Cauchy prior over the standard deviation is
\begin{equation*}
	\lim_{s \to \infty} \left\{ \frac{ \hat{p}_\textrm{\tiny{MML}} } {\hat{p}} \right\} = \frac{n}{n + \alpha} \,,
\end{equation*}
where $\alpha = 1$ in the case of the half-Cauchy prior on the standard deviation and is equal to the prior hyperparameter $\alpha$ in the case of the beta prior. Despite using heavy-tailed prior distributions, the relative bias of the MML estimator for $p$ does not disappear with increasing $s$, unlike in the case of the Poisson distribution with a half-Cauchy prior on the standard deviation (\ref{eqn:poisson:relativebias}). However, as discussed in Section~\ref{ssec:mmlcodelengths}, the use of heavy-tailed priors leads to the same rate of regret growth (i.e., logarithmic in $s$) for both Poisson and geometric distributions. This suggests that comparing Poisson and geometric models on the basis of these codelengths should be largely free of any in-built model selection bias, particularly for large values of $s$.

\section{Simulations}
\label{sec:sim}
We performed simulation experiments to compare the model selection performance of the MML approach with the NML codes and the Bayesian approaches described in Section~\ref{sec:mdl} and Section~\ref{sec:mdl.pois.geom}. In the experiments, we also introduced the known $\mu$ criterion as an ideal reference point. The known $\mu$ criterion compares the negative log-likelihood of the Poisson and geometric models using the actual value of the mean $\mu$
%\paragraph{Known $\mu$ criterion}
%
\begin{align*}
I_{P}(\*x) &= l_{P}(\*x|\mu) \,, \\
I_{G}(\*x) &= l_{G}(\*x|\mu) \,,
\end{align*}
where $\mu$ is the true mean of the data generating model. The known $\mu$ criterion selects the model with the smaller negative log-likelihood as the preferred model for the data. %This method is an ideal reference for the simulation because we do not know $\mu$ in practice.
\subsection{Simulation \RN{1}: Correct model identification}
The aim of this simulation is to discover the frequency of correct hypothesis selection for the MML, MDL and Bayesian model selection techniques. We generated 100,000 data samples from a pre-specified model (Poisson or geometric), with the size of each data sample set to $n=5$, and then calculated the percentage of the correct decisions made by each method. We selected different means for the data generating model and the results for $\mu=2,4,8$ and $80$ are presented in Table \ref{tab:table1}. The percentages of correct detection for the Poisson and geometric models are given in the first two columns, while the third column is the average of the first two columns (i.e., overall frequency of correct detection). For the MML method using conjugate priors (Section \ref{sssec:expbeta}), we set the hyperparameters to $A=5$ for the Poisson model and $\alpha=\beta=1$ for the geometric model (i.e., a uniform prior on $p$). We also used $A=5$ for the MML method with the calibrated conjugate priors (Section \ref{sssec:match1st}). When presenting the results in our simulations, we use ``MML half-Cauchy (s.d)'' and ``MML half-Cauchy (mean)'' to refer to the MML method using a half-Cauchy prior on the standard deviation, and on the square root of the mean respectively (see Section \ref{sssec:cauchy}). For the restricted ANML methods (RANML), we tested three different values for the parameter $\mu^* \in \{10, 100, 1000\}$. 
%Figure~\ref{fig:correct_percent} shows the average percentage of correct decisions made by all the methods tested when $\mu=\{2,\dots,16\}$. 
%
%%!! NEED TO DESCRIBE WHAT IS RANML 10, 100, etc.

\subsection{Simulation \RN{2}: Model selection bias}
In this simulation, we generated 100,000 data samples with sample size $n=5$, with each sample having a $50\%$ probability of being generated from a Poisson model, and a $50\%$ probability of being generated from a geometric model. We expect an unbiased model selection method to select the Poisson (geometric) model for one half of the data samples. The results of this simulation are presented in Table~\ref{tab:table2}. The first two columns show the model selection frequency of the Poisson and geometric models, respectively, for each of the methods considered. The third column is the model selection bias, defined as twice the absolute difference of the observed detection rate for the geometric (Poisson) distribution from 50\% (i.e., an unbiased criterion) for each of the methods considered. A large bias indicates that a criterion has a strong preference for one of the two models.

\begin{table}[h] 
\centering  
\resizebox{\textwidth}{!} {% 
	\begin{tabular}{llllllllll} 
	\toprule 
	\addlinespace[0.2em] 
	& \multicolumn{4}{c}{$\mu = 2$} & 
	& \multicolumn{4}{c}{$\mu = 4$} \\
	\cmidrule(l){2-5} 
	\cmidrule(l){6-10} 
	& Geometric & Poisson & Average & Rank & & Geometric & Poisson & Average & Rank \\
	\cmidrule(l){1-5} 
	\cmidrule(l){6-10} 
	BIC & 54.96 & 89.33 & 72.14 & 11 & & 70.46 & 94.83 & 82.65 & 10 \\
	RANML 10 & 70.75 & 79.91 & 75.33 & 5 & & 77.91 & 91.20 & 84.56 & 5 \\
	RANML 100 & 86.43 & 59.17 & 72.80 & 9 & & 86.40 & 83.90 & 85.15 & 2 \\
	RANML 1000 & 95.99 & 30.46 & 63.22 & 12 & & 93.24 & 69.32 & 81.28 & 12 \\
	ANML two-part & 68.74 & 79.17 & 73.95 & 7 & & 79.57 & 89.20 & 84.39 & 7 \\
	Objective Bayes & 62.40 & 83.21 & 72.81 & 8 & & 77.20 & 89.90 & 83.55 & 9 \\
	Approx Bayes & 86.33 & 58.20 & 72.26 & 10 & & 89.27 & 75.69 & 82.48 & 11 \\
	MML conjugate priors & 73.23 & 80.33 & 76.78 & 2 & & 74.38 & 94.44 & 84.41 & 6 \\
	MML calibrated conjugate & 76.12 & 76.43 & 76.28 & 3 & & 75.94 & 93.92 & 84.93 & 4 \\
	MML half-Cauchy (s.d.) & 77.76 & 74.21 & 75.99 & 4 & & 81.31 & 88.63 & 84.97 & 3 \\
	MML half-Cauchy (mean) & 68.38 & 79.88 & 74.13 & 6 & & 79.45 & 89.07 & 84.26 & 8 \\
	Known mu & 73.69 & 82.04 & 77.87 & 1 & & 85.37 & 90.78 & 88.07 & 1 \\
	\midrule[\heavyrulewidth] 
	\addlinespace[0.2em] 
	& \multicolumn{4}{c}{$\mu = 8$} & 
	& \multicolumn{4}{c}{$\mu = 80$} \\
	\cmidrule(l){2-5} 
	\cmidrule(l){6-10} 
	& Geometric & Poisson & Average & Rank & & Geometric & Poisson & Average & Rank \\
	\cmidrule(l){1-5} 
	\cmidrule(l){6-10} 
	BIC & 83.47 & 98.31 & 90.89 & 11 & & 99.06 & 99.99 & 99.52 & 9 \\
	RANML 10 & 86.51 & 97.33 & 91.92 & 7 & & 99.15 & 99.98 & 99.56 & 8 \\
	RANML 100 & 90.16 & 95.14 & 92.65 & 2 & & 99.27 & 99.96 & 99.62 & 6 \\
	RANML 1000 & 93.84 & 90.09 & 91.97 & 6 & & 99.40 & 99.91 & 99.66 & 2 \\
	ANML two-part & 88.61 & 95.59 & 92.10 & 4 & & 99.39 & 99.88 & 99.63 & 4.5 \\
	Objective Bayes & 87.87 & 95.78 & 91.83 & 8 & & 99.38 & 99.89 & 99.63 & 4.5 \\
	Approx Bayes & 93.70 & 86.92 & 90.31 & 12 & & 99.68 & 98.95 & 99.31 & 10 \\
	MML conjugate priors & 84.02 & 98.35 & 91.19 & 10 & & 99.74 & 15.36 & 57.55 & 12 \\
	MML calibrated conjugate & 84.17 & 98.43 & 91.30 & 9 & & 99.71 & 31.83 & 65.77 & 11 \\
	MML half-Cauchy (s.d.) & 88.07 & 96.50 & 92.28 & 3 & & 99.20 & 99.97 & 99.59 & 7 \\
	MML half-Cauchy (mean) & 88.58 & 95.56 & 92.07 & 5 & & 99.39 & 99.89 & 99.64 & 3 \\
	Known mu & 93.35 & 96.59 & 94.97 & 1 & & 99.86 & 99.96 & 99.91 & 1 \\
	\bottomrule 
	\end{tabular}% 
}
\caption{Percentage of correct identifications of the data generating model (Poisson or geometric) with mean $\mu=\{2,4,8,80\}$ from $100,000$ simulations. The sample size is $n=5$ in each simulation. } 
\label{tab:table1}
\end{table}

\begin{table}[h] 
\centering 
\resizebox{\textwidth}{!} {% 
	\begin{tabular}{llllllllll} 
	\toprule 
	\addlinespace[0.2em] 
	& \multicolumn{4}{c}{$\mu = 2$} & 
	& \multicolumn{4}{c}{$\mu = 4$} \\
	\cmidrule(l){2-5} 
	\cmidrule(l){6-10} 
	& Geometric & Poisson & Bias & Rank & & Geometric & Poisson & Bias & Rank \\
	\cmidrule(l){1-5} 
	\cmidrule(l){6-10} 
	BIC & 32.98 & 67.03 & 34.05 & 11 &   & 38.01 & 61.99 & 23.98 & 11 \\
	RANML 10 & 45.54 & 54.46 &  8.91 & 5 &   & 43.55 & 56.45 & 12.90 & 7 \\
	RANML 100 & 63.54 & 36.46 & 27.07 & 9 &   & 51.39 & 48.61 &  2.79 & 1 \\
	RANML 1000 & 82.79 & 17.21 & 65.58 & 12 &   & 62.11 & 37.89 & 24.22 & 12 \\
	ANML two-part & 44.93 & 55.07 & 10.15 & 6 &   & 45.39 & 54.61 &  9.22 & 4 \\
	Object Bayes & 39.62 & 60.38 & 20.75 & 8 &   & 43.77 & 56.23 & 12.46 & 6 \\
	Approx Bayes & 64.08 & 35.92 & 28.16 & 10 &   & 56.96 & 43.04 & 13.93 & 8 \\
	MML conjugate priors & 46.47 & 53.53 &  7.07 & 3 &   & 40.12 & 59.88 & 19.75 & 10 \\
	MML calibrated conjugate & 49.79 & 50.21 &  0.43 & 1 &   & 41.12 & 58.88 & 17.76 & 9 \\
	MML half-Cauchy (s.d.) & 51.80 & 48.20 &  3.60 & 2 &   & 46.49 & 53.51 &  7.02 & 3 \\
	MML half-Cauchy (mean) & 44.40 & 55.60 & 11.20 & 7 &   & 45.36 & 54.64 &  9.28 & 5 \\
	Known mu & 46.03 & 53.97 &  7.94 & 4 &   & 47.40 & 52.60 &  5.19 & 2 \\
	\midrule[\heavyrulewidth] 
	\addlinespace[0.2em] 
	& \multicolumn{4}{c}{$\mu = 8$} & 
	& \multicolumn{4}{c}{$\mu = 80$} \\
	\cmidrule(l){2-5} 
	\cmidrule(l){6-10} 
	& Geometric & Poisson & Average & Bias &  & Geometric & Poisson & Bias & Rank \\
	\cmidrule(l){1-5} 
	\cmidrule(l){6-10} 
	BIC & 42.58 & 57.42 & 14.84 & 12 &   & 49.88 & 50.12 &  0.25 & 8 \\
	RANML 10 & 44.69 & 55.31 & 10.62 & 9 &   & 49.93 & 50.07 &  0.14 & 3 \\
	RANML 100 & 47.60 & 52.40 &  4.80 & 3 &   & 49.99 & 50.01 &  0.02 & 1 \\
	RANML 1000 & 51.96 & 48.04 &  3.93 & 2 &   & 50.08 & 49.92 &  0.17 & 4 \\
	ANML two-part & 46.56 & 53.44 &  6.87 & 5 &   & 50.09 & 49.91 &  0.18 & 5.5 \\
	Object Bayes & 46.14 & 53.86 &  7.73 & 7 &   & 50.09 & 49.91 &  0.18 & 5.5 \\
	Approx Bayes & 53.51 & 46.49 &  7.03 & 6 &   & 50.70 & 49.30 &  1.40 & 10 \\
	MML conjugate priors & 42.82 & 57.18 & 14.36 & 11 &   & 92.31 &  7.69 & 84.62 & 12 \\
	MML calibrated conjugate & 42.87 & 57.13 & 14.27 & 10 &   & 84.02 & 15.98 & 68.05 & 11 \\
	MML half-Cauchy (s.d.) & 45.89 & 54.11 &  8.22 & 8 &   & 49.95 & 50.05 &  0.09 & 2 \\
	MML half-Cauchy (mean) & 46.59 & 53.41 &  6.81 & 4 &   & 50.09 & 49.91 &  0.19 & 7 \\
	Known mu & 48.40 & 51.60 &  3.20 & 1 &   & 50.29 & 49.71 &  0.58 & 9 \\
	\bottomrule 
	\end{tabular}% 
}
\caption{Model selection bias estimated from $100,000$ simulations. In each simulation, the data has a $50\%$ probability of being generated from a Poisson model, and a $50\%$ probability of being generated from a geometric model, with mean $\mu=\{2,4,8,80\}$. The sample size is $n=5$ in each simulation.} 
\label{tab:table2} 
\end{table}

%\begin{figure}[h]
%\centering
%\includegraphics[width=\textwidth]{correct_percent.pdf}
%\caption{Average percentage of correct identification of the generating model from $100,000$ simulations. The sample size is $n=5$ in each simulation. The generating model mean is $\mu=\{2,4,\dots,16\}$.}
%\label{fig:correct_percent}
%\end{figure}

%\begin{figure}[h]
%\centering
%\includegraphics[width=\textwidth]{regret.pdf}
%\caption{Plot of regret against sufficient statistic in the Poisson model}
%\label{fig:regret}
%\end{figure}

%
%
%
%
\subsection{Discussion of results}
\subsubsection{NML and Objective Bayesian techniques}
\begin{figure}[htbp]
\centering
\includegraphics[width=\textwidth]{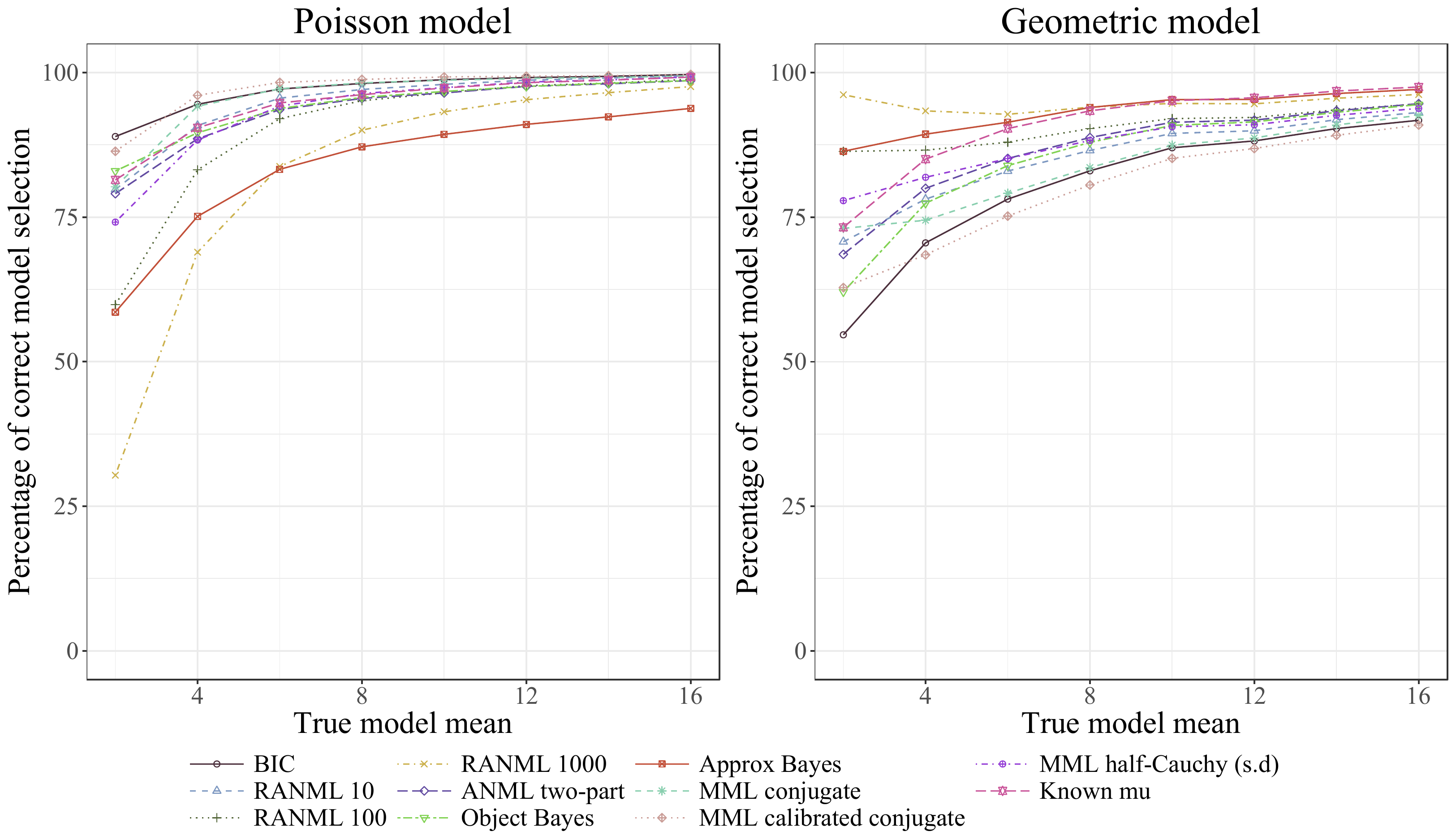}
\caption{Percentage of correct identifications of the Poisson model (left) and geometric model (right) in $100,000$ simulations. The sample size is $n=5$ in each simulation. The generating model mean is $\mu=\{2,4,\dots,16\}$.}
\label{fig:correct_percent2}
\end{figure}

The results for the NML code and the objective Bayesian approaches have been summarized by~\cite{de2006empirical}. Similar results to those observed in~\cite{de2006empirical} were also found in our simulations. BIC performed the worst in terms of percentage of correct detections and model selection bias. As shown in Table~\ref{tab:table2} and Figure~\ref{fig:correct_percent2}, BIC exhibits a strong bias in favor of the Poisson model, particularly when the true mean $\mu$ is small. For example, BIC selected the Poisson model in $67\%$ and $62\%$ of the samples for $\mu=2$ and $\mu=4$, respectively. As expected, the known $\mu$ criterion has the best performance in both tests. All three restricted ANML methods have inconsistent performance for the values of $\mu$ tested; the criteria work very well for one $\mu$ and poorly for another. This is because RANML requires the restricted parameter region to match the range of the data values in order to have good performance. The performance of the two-part restricted ANML and objective Bayesian approaches is robust in both tests, with the two-part restricted ANML having the overall best performance, followed by the objective Bayesian and the approximate objective Bayesian methods.

\subsubsection{MML approaches}
MML with prior distributions \RN{1} (conjugate priors) and \RN{2} (calibrated conjugate priors) exhibits inconsistent results that are similar in behavior to the restricted ANML techniques. MML with priors~\RN{1} and \RN{2} performs well when $\mu=2$ but performs worse as the parameter $\mu$ is increased. From Table~\ref{tab:table1}, we see that these two MML methods are unreliable when $\mu=80$, with $15\%$ and $32\%$ correct detection rates of the Poisson model, respectively. In general, we expect the percentage of correct detection to increase as the sample size $n$ increases or the data generating mean $\mu$ increases, which is not the case for these two methods. The poor performance is related to the selection of the hyperparameter $A$. As discussed in Section~\ref{ssec:remark.prior}, the encoding for the Poisson model is inefficient when $A$ is far from the actual data values. This excess codelength causes the two MML methods to incorrectly select the geometric model. Recall that the hyperparameter is fixed at $A=5$ throughout the simulations which is clearly different from the data generated by a Poisson model with a mean equal to $80$. 
%Figure~\ref{fig:regret} shows a plot of the regret against $(\sum_{i=1}^n x_i)$ in the Poisson model. We can see how fast the regret of the MML with an exponential prior diverges as $s$ increases, compared with a much slower rate for the MML with a Cauchy prior.

MML with half-Cauchy priors performs well in all simulation experiments. In terms of classification bias, MML with a half-Cauchy prior on the standard deviation has excellent performance for $\mu=2$ and $\mu=4$ but slightly favors the Poisson model when $\mu=8$. MML with a half-Cauchy prior on the square root of the mean performs sightly worse and is virtually indistinguishable from the restricted two-part ANML code.

Overall, MML with half-Cauchy priors and the restricted two-part ANML have the best performance in our simulations, with MML half-Cauchy (s.d.) having a slight advantage in terms of correct detection, and the two-part ANML and MML half-Cauchy (mean) having a slight advantage in terms of classification bias when $\mu=8$. These results are not unexpected given the similarity in terms of codelength behavior between the MML half-Cauchy and the restricted two-part ANML code (see Section~\ref{ssec:remark.prior}).

\subsection{Strategy for selecting a prior distribution}
The main conclusion we can draw from the simulations is that finite-sample performance of Bayesian model selection methods can be highly sensitive to the choice of the prior distribution, even in the simple setting studied in this paper. %
%that calibrating priors based on information theoretic properties is a potential strategy for selecting good semi-objective prior distributions. 
%While it appears reasonable to calibrate priors in terms of their prior moments, 
The simulation results show that the strategy of matching prior distributions on their first moments does not necessarily lead to good performance, even in the case of models with one free parameter. Using a highly informative prior such as the exponential distribution (used in prior \RN{2}) with a known prior mean can lead to poor performance if the data is at odds with the prior information, as demonstrated by our experiments. 

On the other hand, using heavy-tailed prior distributions (i.e., half-Cauchy priors \RN{3}) leads to efficient codelengths and good model selection performance regardless of the mean of the data generating process. The simulation results in this paper suggest that heavy tailed priors are a good default choice for modeling location and scale parameters. If subjective information is available, this can be incorporated into heavy tailed priors, such as the half-Cauchy, through hyperparameters (for example, the location and scale hyperparameters). However, in contrast to prior distributions with light tails (i.e., highly informative priors), the codelengths based on heavy tailed priors, and therefore the inferences, are robust to situations in which the prior information is in conflict with the observed data. Therefore, the use of heavy tailed priors appears to strike a good balance between subjectivity and objectivity.

In the case of the Poisson and geometric problem studied in this paper, the half-Cauchy priors result in MML codelengths for the two models that have the same (logarithmic) growth of regret as the sufficient statistic $s \to \infty$. Calibrating codelengths in this manner appears to be a parameterization independent approach to specifying priors when there exists no subjective information. The performance of such regret-calibrated priors in more general settings is an interesting topic for future work.

%In general, if we do not have known prior information, using a vague prior like heavy tailed prior is a good strategy. 
%  Furthermore, we can also put subjectivity on the heavy-tailed prior, for example, by modifying the scale in a half-Cauchy prior or the degrees of freedom in a $t-$distribution prior.

%Figure~\ref{fig:regret} shows a plot of the regret against $(\sum_{i=1}^n x_i)$ in the Poisson model for the MML and two-part ANML techniques. We see that the regret of MML with an exponential prior diverges as $s$ increases, compared with a much slower divergence rate for MML with a Cauchy prior. Furthermore, we see that the rate of the regret for the MML and two-part ANML methods is similar with MML Cauchy having a smooth curve as opposed to the jump discontinuities observed in the two-part ANML approach.
%
%
\section{Conclusion}
\label{sec:con}
This paper has presented several MML approaches to the model selection problem involving data generated from a Poisson or a geometric distribution. In the MML approach, prior distributions for the parameters are required and we have proposed three candidate prior densities: (\RN{1}) conjugate priors, (\RN{2}) calibrated conjugate priors, and (\RN{3}) a half-Cauchy prior on either the standard deviation or the square root of the mean of both models. These three MML methods are then compared to MDL approaches based on the NML code and the objective Bayesian approaches presented by~\cite{de2006empirical}. We compared the performance of these methods in terms of the percentage of making a correct detection and the classification bias.

We found that using a half-Cauchy prior leads to good model selection results for the MML criterion. In particular, we found that a half-Cauchy prior on the standard deviation performed slightly better than a half-Cauchy prior on the square root of the mean. In contrast, using a conjugate exponential prior with an \emph{a priori} selected hyperparameter for the Poisson model leads to poor performance if the choice of hyperparameter is in conflict with the data. We also found that the restricted two-part ANML criterion has similar performance to the MML code based on the half-Cauchy prior. Most of the methods tested have excellent performance if the mean of the data generating model is moderate to large. Overall, we recommend using the MML criterion with a half-Cauchy prior on the standard deviation when comparing Poisson and geometric models. The results of our simulations suggest that calibrating heavy-tailed prior distributions based on their asymptotic rate of regret is a promising approach for specifying priors in non-nested model selection problems.
%Although this is a relatively simple model selection problem, we can get an idea of the properties of MML and NML under various sample sizes in a more general setting. 
%The results in this paper provide further insight for researchers when deciding which model selection method to apply in practice.

%\clearpage
\bibliographystyle{agsm}
\bibliography{mybibfile}

\end{document}